\documentclass{article}[12pt,a4paper,fullpage]   
\usepackage[left=3cm, right=3cm, top=3cm, bottom=4cm]{geometry}
\usepackage{amsmath}
\usepackage{amssymb}
\usepackage{cite}
\usepackage{url}
\usepackage{graphicx}
\usepackage{enumitem}
\usepackage{array}
\usepackage{hyperref}
\usepackage{color}
\usepackage{multirow}
\usepackage{subcaption}
\usepackage{float}
\usepackage{algorithm}
\usepackage[noend]{algpseudocode}
\usepackage{parskip}
\usepackage{mathtools}
\usepackage{pgfplotstable} 
\usepackage{booktabs}
\usepackage{array,multirow,makecell}
\usepackage{acro}
\DeclareAcronym{AO}{
  short=AO,
  long=Adaptive Optics,
}
\DeclareAcronym{GMT}{
  short=GMT,
  long=Giant Magellan Telescope,
}
\DeclareAcronym{ELT}{
  short=ELT,
  long=Extremely Large Telescope,
}
\DeclareAcronym{ELTs}{
  short=ELTs,
  long=Extremely Large Telescopes,
}
\DeclareAcronym{VCM}{
  short=VCM,
  long=Voice Coil Motor,
}
\DeclareAcronym{FE}{
  short=FE,
  long=Finite Element,
}
\DeclareAcronym{ODEs}{
  short=ODEs,
  long=Ordinary Differential Equations,
}
\DeclareAcronym{LTI}{
  short=LTI,
  long=Linear Time Invariant,
}
\DeclareAcronym{SISO}{
  short=SISO,
  long=Single Input Single Output,
}
\DeclareAcronym{MIMO}{
  short=MIMO,
  long=Multiple Input Multiple Output,
}
\DeclareAcronym{SVD}{
  short=SVD,
  long=Singular Value Decomposition,
}
\DeclareAcronym{IRKA}{
  short=IRKA,
  long= Iterative Rational Krylov Algorithm,
}
\DeclareAcronym{ITIA}{
  short=ITIA,
  long= Iterative Tangential Interpolation Algorithm ,
}
\DeclareAcronym{ISTIA}{
  short=ISTIA,
  long= Iterative SVD Tangential Interpolation Algorithm ,
}
\DeclareAcronym{FISTIA}{
  short=FISTIA,
  long= Frequency Limited Iterative SVD Tangential Interpolation Algorithm ,
}
\DeclareAcronym{RB}{
  short=RB,
  long=Reference Body,
}
\DeclareAcronym{HO}{
  short=HO,
  long=Higher Order,
}
\DeclareAcronym{LF}{
  short=LF,
  long=Loewner Framework,
}
\DeclareAcronym{BT}{
  short=BT,
  long=Balanced Truncation,
}

\newcommand{\norm}[1]{\left\|#1\right\|}

\DeclareMathOperator{\diag}{diag}
\newcolumntype{R}[1]{>{\raggedleft\arraybackslash }b{#1}}
\newcolumntype{L}[1]{>{\raggedright\arraybackslash }b{#1}}
\newcolumntype{C}[1]{>{\centering\arraybackslash }b{#1}}
\begin{document}
\title{High fidelity adaptive mirror simulations\\with reduced order models}
\author{Bernadett Stadler\thanks{RICAM, Altenbergerstra{\ss}e 69, A-4040 Linz, Austria,
  (bernadett.stadler@ricam.oeaw.ac.at, 
  ronny.ramlau@ricam.oeaw.ac.at).}
\and Roberto Biasi~\thanks{Microgate, Via Waltraud Gebert Deeg 3e, 39100, Bolzano, Italy, (roberto.biasi@microgate.it,
mauro.manetti@microgate.it).}
\and Mauro Manetti\footnotemark[2]
\and Andreas Obereder~\thanks{Industrial Mathematics Institute, JKU Linz, Altenberger Str.~69, 4040 Linz, Austria, (andreas.obereder@mathconsult.co.at, ronny.ramlau@jku.at).}
\and Ronny Ramlau\footnotemark[1]\hspace{0.2cm}\footnotemark[3]
\and Matteo Tintori~\thanks{A.D.S. International, Via Pio Galli sindacalista 3, 23841, Annone di Brianza, Italy, (m.tintori@ads-int.com).}
}
\title{High fidelity adaptive mirror simulations\\with reduced order models}

\maketitle

\begin{abstract}
In the design process of large adaptive mirrors numerical simulations represent the first step to evaluate the system design compliance in terms of performance, stability and robustness. For the next generation of Extremely Large Telescopes increased system dimensions and bandwidths lead to the need of modeling not only the deformable mirror alone, but also all the system supporting structure or even the full telescope. The capability to perform the simulations with an acceptable amount of time and computational resources is highly dependent on finding appropriate methods to reduce the size of the resulting dynamic models. In this paper we present a framework developed together with the company Microgate to create a reduced order structural model of a large adaptive mirror as a preprocessing step to the control system simulations. The reduced dynamic model is then combined with the remaining system components allowing to simulate the full adaptive mirror in a computationally efficient way. We analyze the feasibility of our reduced models for Microgate's prototype of the adaptive mirror of the Giant Magellan Telescope.\\


\noindent \textbf{AMS:}
65-04, 74-10, 85-08, 85-10\\

\end{abstract}

\section{Introduction}\label{sect_introduction}
For ground-based telescopes, so-called \ac{AO} systems are used to compensate the image distortions in astronomical observations caused by atmospheric turbulence, using the flexible shape of an optical surface. The control of an AO system is complex and requires at least a wavefront sensor to get information about the atmospheric turbulence causing the image distortions, a known natural or laser guide star as reference source, a mirror shape command generator, and a deformable mirror to compensate for the atmospheric turbulence. 

We consider the control of an adaptive mirror based on non-contacting voice-coil actuators, which are co-located to capacitive position sensors \cite{Biasi2010_2}. This deformable mirror technology was developed by the company Microgate\footnote{\url{https://engineering.microgate.it}} together with other partners (ADS International\footnote{\url{https://www.ads-int.com/}}, INAF-Osservatorio Astrofisico di Arcetri\footnote{\url{https://www.arcetri.inaf.it/en/}}, and the Aerospace Engineering Department of Politecnico di Milano\footnote{\url{https://www.aero.polimi.it/it/il-dipartimento}}). It has been already deployed realizing the adaptive secondary mirror of several large telescopes, including the Multiple Mirror Telescope, the Large Binocular Telescope, and the Very Large Telescope. Microgate is engaged in the final design and construction of the adaptive mirrors for the next generation of \ac{ELTs}. Sub-system and full-system multiphysics simulation plays a key role in the design phase of such complex projects. Therefore, numerical simulation has always been intensively pursued by the company. Several design solutions need to be compared and the impact of different modelling choices have to be verified. 

The structural dynamic models used to describe the large adaptive mirror are of high order, caused by the need to describe well high spatial order deformations and by the inherent complexity of the system supporting structure. Hence, a reduced order mirror model is required that guarantees high fidelity results with a reasonable simulation time. The earliest methods for model order reduction techniques go back to the 1960s in the field of structural dynamics. These methods rely on the identification of eigenfrequencies and are referred to as mode displacement methods \cite{Rayleigh1945,Geradin1997}. In the 1980s the important reduced order method balanced truncation \cite{Moore1981,Enns1984} has been developed in the system and control theory community. In the field of numerical mathematics, approaches such as Pade-via-Lanczos and rational interpolation methods came up in the 1990s and are still under research, e.g. in \cite{Feldmann1995,Grimme1997,Antoulas2020}. Nowadays these methods are often used for the design and analysis of large electronic circuits. In recent years data driven approaches gained a lot of traction, see e.g. \cite{Gruyter2021,Loewner2022,Gosea2022,Kutz2016,Antoulas1986,Antoulas2020,Gosea2021}. In the literature various reviews on model order reduction techniques exist \cite{Craig1968,Klerk208,Gugercin2004,Bai2002,Freund2003,Antoulas2005,Antoulas20052,Gallivan1999,Besselink2013,Benner2015,Benner2017,Beattie2022}. 

In the framework of this paper a reduced order model is created by applying different methods as a preprocessing step to the control system simulations. In \cite{Manetti2010} the authors used balanced truncation for reducing the complexity of an adaptive mirror model, which is a common approach for control theory applications. However, in contrast to the model presented in this paper theirs did not take into account the system supporting structure up to the full telescope and was thus less complex. Moreover, here we also consider Krylov subspace based methods and the Loewner framework. We combine the reduced order structural model with the remaining system components and run full system simulations using different numerical tools. Our developments are validated via simulations of Microgate's P72 \cite{Gallieni2020}, which is a $72$ actuation points prototype of the \ac{GMT}.

The outline of this paper is as follows: We start with a brief description of the physical model of the adaptive mirror in Section~\ref{sect:phys_model}. Section~\ref{sec:mor} is dedicated to an overview of existing model order reduction methods from the field of structural dynamics, control theory, and numerical mathematics. In Section~\ref{sec:digital_twin} we describe our framework for performing high fidelity mirror simulations with a reduced order structural model and we give some implementation details. Numerical simulations for the GMT P72 adaptive mirror, including a performance evaluation of different reduced order methods, are shown in Section~\ref{sec:numerical_simuls}. We end with a summary and conclusions in Section~\ref{sec:conclusion}.

\section{Physical model}\label{sect:phys_model}
\begin{figure}[ht]
\centering
  \includegraphics[width=0.6\textwidth]{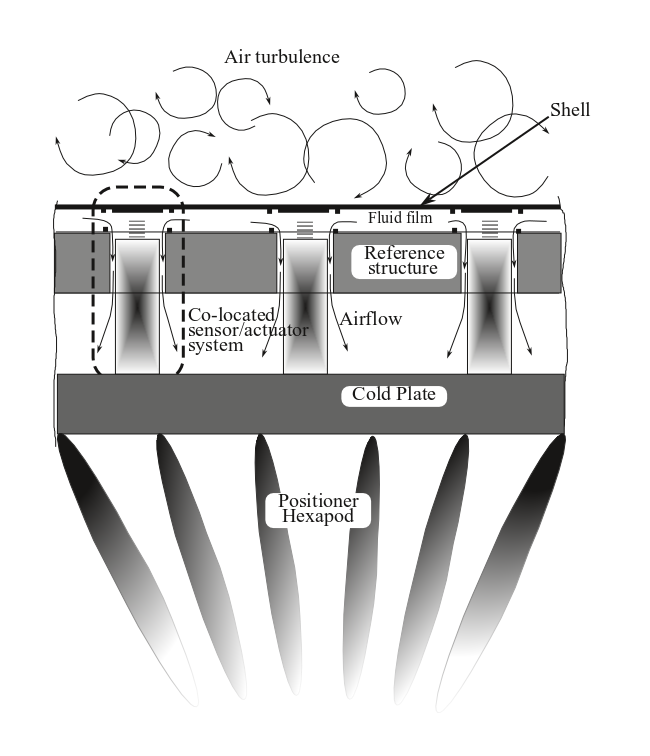}
  \caption{Deformable mirror and supporting structure.}
  \label{fig:mirrorSketch}
\end{figure}

The physical modeling of the deformable mirrors treated in this paper usually requires a multiphysics description. The level of modeling accuracy and complexity can be set on the basis of the needs and usually involves the description of (see Figure~\ref{fig:mirrorSketch}): 
\begin{itemize}
    \item a thin deformable mirror, usually a Zerodur shell about 2 mm thick;
    \item a reference structure called reference body, also made of Zerodur to grant the required thermal stability;
    \item the squeeze film action of the air trapped in the thin gap, about 100 microns, between the mirror and the reference body, which significantly contributes to the deformable mirror damping, affecting both the control loop stability and the controlled system performance;
    \item the cold plate, typically made by aluminum, where the \ac{VCM} actuators are mounted, which sustains the VCM reaction forces providing adequate mechanical stability and taking care of the system cooling;
    \item the system positioner, typically a hexapod, responsible for the system alignment and offloading of the mirror low-order modes;
    \item the mirror capacitive sensors and VCM actuators with their signal conditioning and the digital feedforward-feedback control loop.
\end{itemize}

The focus of this paper is on model order reduction methods for the structural dynamics of the system components mentioned above, even if the effectiveness of the reduced order models are verified also through the simulation of the full system behavior. Details about the fluid dynamic model and the control system are omitted and can be found in \cite{Manetti2012}.

The deformable mirror is described using a linear structural model in the time domain. One way to represent the structural dynamics is via its degrees of freedoms using second order differential equations. Another way is given by its system states with first order differential equations often referred to as state space representation. This formulation is preferred by control engineers as linear control system analyses and design methods are usually given in the state space form. Besides the form of equations the model can be represented via different coordinate systems. Commonly, nodal and modal coordinates are employed. The nodal coordinates are expressed through displacements and velocities of specific physical locations, also called nodes. In contrast, modal coordinates are defined through structural eigenmodes \cite{Gawronski2004}.

\subsection{Second order structural models}
The structural model of the adaptive mirror shape can be represented by second order linear differential equations using nodal coordinates, i.e., in terms of displacement, velocity and acceleration. Typically, the \ac{FE} method with piecewise polynomial basis functions \cite{Zienkiewicz2005} is applied, where the structure is described by the mass, stiffness and damping matrices. The discretized displacement $u_h$, which depends on the spatial variable $z$ and the time $t$, is represented via
\begin{equation}\label{eq:femiso}
   	u_h(z,t)=\sum_{i=1}^{n_g} \psi_i(z)u_i(t),
\end{equation}
where $n_g$ is the number of degrees of freedom, $\psi_i(z)$ are the linear independent FE basis functions, and $(u_i)_{i=1}^{n_g} \eqqcolon \vec{u}_h(t) \in \mathbb{R}^{n_g}$ are the time dependent coefficients of the FE solution $u_h$. With Equation~\eqref{eq:femiso} we can identify the FE function $u_h$ with its coefficient vector $\vec{u}_h$. Assuming a linear structural model described by its degrees of freedom the FE discretization leads to the following system of \ac{ODEs}
\begin{equation}\label{eq:structural}
    M_g\ddot{\vec{u}}_h+D_g\dot{\vec{u}}_h+K_g\vec{u}_h=B_g(f_a^c+f_a^d),
\end{equation}
where $M_g$ is the $n_g\times n_g$ mass matrix, $D_g$ is the $n_g\times n_g$ damping matrix, $K_g$ is the $n_g\times n_g$ stiffness matrix and $B_g$ is the $n_g\times n_a$ force influence matrix. The control and disturbance force vectors at the $n_a$ actuation points are denoted by $f_a^c$ and $f_a^d$, respectively. For more details about the mechanical model and the coupling with the fluid part we refer to \cite{Manetti2010}.

Equation \eqref{eq:structural} can be formulated in modal coordinates. When considering free vibrations of a structure without damping, i.e., a structure without external excitation and with a damping matrix $D=0$, the vibration modes are found using a time harmonic representation of the displacement of the unforced system in \eqref{eq:structural}. This leads to the following generalized eigenvalue problem
\begin{equation}
    (K_g-\omega_i^2M_g)\phi_i=0,
\end{equation}
where $\omega_i^2$ with $i\in \{1,\dots,n_m\}$ is an eigenfrequency and $\phi_i$ the corresponding mode shape vector. The so-called modal displacement can be written as
\begin{equation}\label{eq:modalDispl}
    q_m = \sum_{i=1}^{n_m} \phi_i q_i,
\end{equation}
with the modal coefficients denoted by $q_i$. The displacement written in matrix notation becomes
\begin{equation}\label{eq:matrixNodal}
    q_m=\Phi_g q, \quad \Phi_g=[\phi_1, \phi_2, \dots, \phi_{n_m}] \in \mathbb{R}^{n_g \times n_m},
\end{equation}
with the $n_g\times n_m$ modal shape matrix $\Phi_g$ normalized such that
\begin{align}
    M_m & \coloneqq \Phi_g^TM_g\Phi_g = I,\\
    K_m & \coloneqq \Phi_g^TK_g\Phi_g = \diag(\omega_i^2).
\end{align}
The adaptive mirrors considered in this paper experiences, at least most of the time, only light damping. In this setting it is common, see e.g. \cite{Gawronski2004}, to  include a diagonal modal damping approximation with damping coefficients $\xi_i$
\begin{equation}
         D_m \coloneqq \diag\{2\xi_i\omega_i\} \approx \Phi_g D_g \Phi_g^T.
\end{equation}
The force influence matrix is transformed similarly
\begin{align}
    B_m \coloneqq \Phi_g^T B_g.
\end{align}
In this way the equation of motion \eqref{eq:structural} in modal coordinates becomes fully uncoupled 
\begin{equation}
    \ddot{q_i}+2\xi_i\omega_i\dot{q}_i+\omega^2_iq_i = f_{m_i}^c+f_{m_i}^d, \quad i\in\{1,\dots,n_m\},
\end{equation}
where $f_{m_i}^c$ and $f_{m_i}^d$ denote the $i$-th component of the modal force vectors
\begin{align}
    f_{m}^c \coloneqq B_mf_a^c,\\
    f_{m}^d \coloneqq B_mf_a^d.
\end{align}

Note that for the numerical simulations we focus on modal coordinates as the FE model provided by the company is given in this representation.

\subsection{State space representation}
For control theory applications it is common to rewrite the second order differential equations in \eqref{eq:structural} into a so-called state space representation, which is a set of first order differential equations. We consider the \ac{LTI} state space representation of the form
\begin{align}
\dot{x} & = Ax+Bu\label{eq:LTI1}\\
y & = Cx\label{eq:LTI2},
\end{align}
where $u\in\mathbb{R}^{m}$ is the input and $y\in \mathbb{R}^{p}$ the output of the system. The state vector is denoted by $x\in\mathbb{R}^n$ and $A\in\mathbb{R}^{n\times n}$ is commonly referred to as state matrix. The input matrix is denoted by $B\in\mathbb{R}^{n\times m}$ and the output matrix by $C\in\mathbb{R}^{p\times n}$.

By applying the Laplace transform the model can be represented via its transfer function, which describes the input-output behavior of the system. In this paper we assume to have a stable system, i.e., all eigenvalues of $A$ have a negative real part \cite{Gawronski2004}. Moreover, we assume a minimal realization, i.e., all components of the state vector contribute to the input-output behavior. The transfer function of the system in \eqref{eq:LTI1}-\eqref{eq:LTI2}  is given by
\begin{equation}\label{eq:transferFunc}
    H(s) = C(sI-A)^{-1}B, \quad s\in\mathbb{C}.
\end{equation}
Note that the transfer function is invariant under coordinate transformation.

The modal version of the state space representation \eqref{eq:LTI1}-\eqref{eq:LTI2} of flexible structures has a special form, in which the state matrix $A_m$ in modal coordinates is block diagonal
\begin{align*}
    A_m=\diag(A_{m_i}),
\end{align*}
with $A_{m_i}$ being $2\times 2$ blocks, see e.g.  \cite{Gawronski2004}. The input and output matrices in modal coordinates are arranged accordingly
\begin{equation}
    B_m=[B_{m_1},\dots,B_{m_{n_m}}]^T \quad C_m=[C_{m_1},\dots,C_{m_{n_m}}].
\end{equation}
The state vector is split into $n_m$ components that represent the state of a certain mode
\begin{equation}
    x=[x_1,\dots,x_{n_m}]^T, \quad x_i=[x_{i_1},x_{i_2}].
\end{equation}
The $i$-th component, i.e., the $i$-th mode, has the state representation $(A_{m_i},B_{m_i}, C_{m_i})$.

We consider the following representation for the blocks of the state space matrices
\begin{equation}\label{eq:modalLTI}
    A_{m_i}=
    \begin{bmatrix}
        0 & 1 \\
        \omega_i & -2\xi_i\omega_i
    \end{bmatrix},
    \quad 
    B_{m_i} = 
    \begin{bmatrix}
         0  \\
         b_{m_i} 
    \end{bmatrix},
    \quad 
    C_{m_i} = 
    \begin{bmatrix}
         0  \\
         c_{m_i} 
    \end{bmatrix}.
\end{equation}
The $i$-th state is given by
\begin{equation}
    x_i=[q_{m_i}, \dot{q}_{m_i}]^T,
\end{equation}
where $q_{m_i}$ is the $i$-th modal displacement and $\dot{q}_{m_i}$ the modal velocity. Each component has a modal displacement and velocity, which are related to the original one by \eqref{eq:modalDispl}. Note that this leads also to a special form of the transfer function in modal coordinates which we denote by $ H_m(s)$. For more details we refer to \cite{Gawronski2004}.

\section{Model order reduction}\label{sec:mor}
In practice, we deal with a very large number of degrees of freedom for the structural model required to obtain a certain modeling accuracy. Moreover, for the design and construction of the large adaptive mirrors a high number of simulations has to be performed. This makes the direct use of a FE model infeasible and a reduced order model, which represents well the dynamics of the structure within the frequency band of interest, has to be used.

For the control of an adaptive mirror it is important that the reduced order model preserve the input-output behavior. Thus, the quality of the reduced model can be evaluated by comparing its output to the output of the high order model. This is commonly done by considering the error in the transfer function. The $\mathcal{H}_2$ error of a stable system is defined by
\begin{equation}\label{eq:errorTF2}
    \norm{H_r-H}_{\mathcal{H}_2}, \quad \norm{H}_{\mathcal{H}_2}=\sqrt{\frac{1}{2\pi}\int_{-\infty}^\infty \text{tr}\{H^T(-j\omega)H(j\omega)\}d\omega},
\end{equation}
where we denote by $j$ the imaginary unit and $H$ and $H_r$ are the transfer functions of the high and reduced order model, respectively. Attention is also paid to the $\mathcal{H}_\infty$ error
\begin{equation}\label{eq:errorTFInf}
    \norm{H_r-H}_{\mathcal{H}_\infty}, \quad \norm{H}_{\mathcal{H}_\infty}=\text{sup}_{\omega\in \mathbb{R}} \sigma_{\text{max}}(H(j\omega)),
\end{equation}
where $\sigma_{\text{max}}(.)$ is the largest singular value. There exist several algorithms that aim for minimizing the $\mathcal{H}_2$ error, i.e., the so-called first order optimality condition,
\begin{equation}\label{eq:firstOpt}
     \norm{H_r-H}_{\mathcal{H}_2} \rightarrow \text{min} \text{, for stable } H_r.
\end{equation}

In the following we summarize model order reduction approaches based on modal truncation, moment matching and data-driven interpolation.

\subsection{Modal truncation}\label{sec:modalTruncatin}
The modal displacement method \cite{Geradin1997} computes a reduced model based on free vibration modes of the structure. For an acceptable approximation of the dynamics of the mirror a relative small number of modes may be sufficient. To this end not all $n_m$ eigenvectors in \eqref{eq:modalDispl} are used, but a limited number $r<n_m$. Usually, one is interested in the response of the system for lower frequencies, as most structural forcing terms operate at low frequencies. Hence, the first $r$ eigenvectors are kept, which correspond to the $r$ lowest eigenfrequencies, and the others are truncated. Note that to accurately control complex mirror deformations a number of modes at least equal to the number of actuators is required to correctly represent the number of independent degrees of freedom of the control problem.

\subsection{Balanced truncation}
Balanced truncation \cite{Moore1981,Enns1984} is a very popular model order reduction method which allows to reach the first order optimality condition in \eqref{eq:firstOpt}. The LTI system is transformed into a so-called balanced realization, in which the states are ordered according to their contribution to the input-output behavior. To quantify the systems input-output behavior the notions of controllability and observability are introduced. These system properties are determined via the so-called controllability $W_c\in\mathbb{R}^{n\times n}$ and observability $W_o\in\mathbb{R}^{n\times n}$ Gramians \cite{Moore1981}. We are interested in the stationary or time invariant solution. In this case the controllability and observability Gramians, see e.g. \cite{Zhou1996}, can be obtained by solving the Lyapunov equations
\begin{align}\label{eq:Lyapunov}
    AW_c+W_cA^T+BB^T=0,\\
    A^TW_o+W_oA+C^TC=0,
\end{align}
where $A\in\mathbb{R}^{n\times n}$, $B\in\mathbb{R}^{n\times m}$ and $C\in\mathbb{R}^{p\times n}$. The Hankel singular values $\gamma_i$ with $i=1,\dots,n$, used to determine the importance of a certain state for the input-output behavior, are defined as the square roots of the eigenvalues of the product $W_cW_o$. Note that in contrast to Gramians the Hankel singular values are system invariants. Once the system is transformed into the balanced representation the states with small influence on the system dynamics, i.e., with small Hankel singular values, are discarded. The reduced system is again in balanced form, therefore, it preserves the stability property of the original system \cite{Pernebo1982}. This is a very important point for adaptive mirror control. Moreover, an upper bound on the error \cite{Enns1984,Glover1984} is given by
\begin{equation}
    \norm{H_r(s)-H(s)}_{\mathcal{H}_\infty} \leq 2\sum_{i=r+1}^n \gamma_i,
\end{equation}
i.e., we obtain a good approximation as long as the truncated Hankel singular values are small.

In balanced truncation the total frequency range is covered, however, for adaptive mirror control there is a certain frequency range of interest. For the frequency limited balanced truncation, the controllability and observability Gramians are computed in finite frequency intervals. We refer to \cite{Gawronski1990} for more details on the computation. Similar extensions exist to, e.g., time-limited balanced truncation or frequency weighted balanced truncation, e.g. in \cite{Gugercin2004}, but are not considered in this paper.

\subsubsection{Modal approximation}
A huge drawback of balanced truncation is that solving the Lyapunov equations in \eqref{eq:Lyapunov} is computationally very demanding. Thus, applying balanced truncation in this form would not be feasible for our large scale system. However, the representation of the state space in modal form \eqref{eq:modalLTI} leads to a special form of the controllability and observability Gramians \cite{Gawronski2004}. Assuming small damping the Gramians can be approximated via
\begin{align}
    W_c \approx \diag(w_{c_i}I), \quad w_{c_i} = \frac{\norm{B_{m_i}}_2^2}{4\xi_i\omega_i},\label{eq:GramiansModal1}\\
    W_o \approx \diag(w_{o_i}I), \quad w_{o_i} = \frac{\norm{C_{m_i}}_2^2}{4\xi_i\omega_i},\label{eq:GramiansModal2}
\end{align}
where $w_{c_i}$ and $w_{o_i}$ are called modal controllability and observability factors. The approximate Hankel singular values in modal coordinates are the geometric means of these two factors
\begin{equation}\label{eq:HankelModal}
    \gamma_i \approx \sqrt{w_{c_{ii}}w_{o_{ii}}}.
\end{equation}
Equations \eqref{eq:GramiansModal1}-\eqref{eq:GramiansModal2} can be evaluated very fast. Moreover, the system preserves its original modal representation.

More recently also a data-driven approach of balanced truncation has been developed in \cite{Gosea2022} in which the product of the Gramians is approximated directly through frequency response data up to a desired accuracy.

\subsection{Moment matching methods}
Similar to balanced truncation, the goal of moment matching methods is to reduce the system to one with fewer degrees of freedom, but with similar input-output behavior, i.e., approximating the transfer function. Typically, those methods are used for large electronic systems, but also for structural vibrations, e.g., in \cite{HawLing2010}. The basic approach is to approximate the state space $S$ of $x$ by a low dimensional subspace $\tilde{S}$ via projection.

For an arbitrary interpolation point $s_0\in\mathbb{C}$ the transfer function can be written via its moment expansion
\begin{align}\label{eq:Taylor}
    H(s)=\sum_{i=0}^\infty (-1)^iM_i(s_0)(s-s_0)^i,\\
    M_i(s_0)\coloneqq C^T(s_0I-A)^{-i}(sI-A)^{-1}B.
\end{align}
where the $M_i$ are the so-called moments. Note that for Multiple Input Multiple Output (MIMO) systems the moments are $p\times m$ matrices. The idea behind moment matching methods is to truncate the Taylor series in \eqref{eq:Taylor} and thus find a reduced order transfer function $H_r$, which matches the first $r$ moments of $H$ at the given expansion point $s_0$. The precision of the moment matching methods depends not only on the number of moments matched but also highly depends on the chosen expansion point $s_0$. The reason is that the Taylor series in \eqref{eq:Taylor} is only a reasonable approximation within a certain distance from $s_0$. To increase accuracy methods that use multiple expansion points $s_1\dots,s_\ell\in\mathbb{C}$ are often used. In the SISO case, the moment matching problem is given by finding an $H_r$ such that
\begin{equation}\label{eq:SISOMM}
    H_r(s_i) = H(s_i) \quad \text{ for } i=1,\dots,N,
\end{equation}
which is also known as rational interpolation. In the MIMO case the moments are matrices and one commonly interpolates along certain directions, which leads to the so-called tangential interpolation problem, see e.g. \cite{VANDOOREN2008}. The aim is to find a rational matrix $H_r$ for given interpolation points $\lambda_i\in\mathbb{C}$ and $\mu_i\in\mathbb{C}$ and the corresponding right and left tangential directions $r_i\in\mathbb{C}^m$ and $\ell_i\in\mathbb{C}^p$ such that 
\begin{align}\label{eq:MIMOMM}
    H_r(\lambda_i)r_i=H(\lambda_i)r_i \quad \text{ for } i=1,\dots,k,\\
    \ell_j^TH_r(\mu_j)=\ell_j^TH(\mu_j) \quad \text{ for } j=1,\dots,q.
\end{align}
The interpolation points and tangential directions have to be selected in advance to realize certain model reduction goals.

\subsubsection{Rational Krylov subspace methods}
A common way to deal with the moment matching problem is rational interpolation by projection. Here the moments are used to construct the biorthogonal projection matrices $V,W\in\mathbb{R}^{n\times r}$, with which the reduced order model is obtained via Petrov-Galerkin projection
\begin{align}\label{eq:Petrov-Galerkin}
        A_r = W^TAV,\quad B_r&=W^TB, \quad C_r=C^TV,\\
        x &\approx Vx_r,
\end{align}
where $A_r\in\mathbb{R}^{r\times r}$, $B_r\in\mathbb{R}^{r\times m}$, $C_r\in\mathbb{R}^{p\times r}$ and $x_r\in\mathbb{R}^{r}$. The columns of $V$ form an orthonormal basis of a subspace $\tilde{S}\subset S$ of the space of the state vector. Assuming the SISO case with a single expansion point $s_0\in\mathbb{C}$, $\tilde{S}$ is spanned by
\begin{align}\label{eq:spanV}
        \text{range}\{V\} = \text{span}\{(s_0I-A)^{-1}B,(s_0I-A)^{-1}(s_0I-A)^{-1}B,
        \dots,\\\nonumber((s_0I-A)^{-1})^{r-1}(s_0I-A)^{-1}B\}.
\end{align}
For the right projection matrix $W$ a similar property holds
\begin{align}\label{eq:spanW}
        \text{range}\{W\} = \text{span}\{(s_0I-A)^{-T}C^T,(s_0I-A)^{-T}(s_0I-A)^{-T}C^T,
        \dots,\\\nonumber((s_0I-A)^{-T})^{r-1}(s_0I-A)^{-1}C^T\}.
\end{align}
The range of $V$ and the range of $W$ span so-called Krylov subspaces. Note that when multiple expansion points $s_1\dots,s_\ell\in\mathbb{C}$ are used, the Krylov subspaces get extended. In the SISO case $B$ and $C$ are vectors, thus, for the construction of $V$ and $W$ rational Krylov subspace algorithms, such as rational Lanczos or rational Arnoldi, can be used \cite{Grimme1997}. These methods are very efficient as only matrix factorization, forward and backward substitution as well as matrix-vector multiplications are involved. Their  complexity is of $\mathcal{O}(nr^2)$. For adaptive mirror simulations we are dealing with the MIMO case, where $B$ and $C$ are matrices and the tangential directions need to be included into Krylov subspaces. Here, block rational Arnoldi or Lanczos methods can be employed. 

Recently methods that start from an initial group of interpolation points and iteratively update them have been developed. One of those methods is the \ac{IRKA}. Initially this algorithm was developed for SISO systems in \cite{Gugercin2008}. In \cite{VANDOOREN2008} it was extended to the MIMO case called the \ac{ITIA}. This method creates a reduced order model that fulfills the first order $\mathcal{H}_2$ optimality condition in \eqref{eq:firstOpt}. The updated expansion points are selected as the negative poles of the transfer function of the reduced model $H_r$ or equivalently as the negative eigenvalues of the state matrix\cite{VANDOOREN2008}.

In contrast to balanced truncation, moment matching methods, in general, do not preserve the stability of the system. For systems with special structures there exist approaches where the reduced order model is guaranteed to be stable, see \cite{Odabasioglu1998}. In \cite{Vassal2011} the authors proposed the \ac{ISTIA}, which combines balanced truncation and Krylov subspace based methods. For one of the projection matrices the observability or controllability Gramian is used, whereas the other projector is computed using Krylov subspace based methods. The authors show that due to the usage of a Gramian preserves stability. When frequency limited controllability or observability Gramians are used, the method is commonly referred to as \ac{FISTIA} \cite{Vassal2011}.

An advantage of the Krylov subspace based methods it that their efficiency does not rely on a modal representation, but only depends on the system size. The more moment vectors are included into the subspace, the more accurate the approximation is, however, the reduced order model is enlarged. In practice, a trade-off between accuracy and the reduction size has to be made. If both projection matrices are constructed via Krylov subspace methods it is commonly referred to as two sided, whereas if only one of the projection matrices is constructed in this way it is called one sided. Using the two sided method results in $2r$ matched moments, whereas the one sided method matches $r$ moments \cite{Benner2021}.

\subsubsection{Loewner framework}
The Loewner framework is a data-driven interpolation method for which the frequency response data, i.e., samples of the transfer function $H$ at certain frequency points $s$, is used. In the MIMO case, where the moments are $p\times m$ matrices, the tangential directions are included into the data set and the data is split into a right part
\begin{align}\label{eq:rightData}
    M=\diag(\mu_1,\dots,\mu_q)\in\mathbb{C}^{q\times q},\quad L=[\ell_1^T,\dots,\ell_q^T]^T\in\mathbb{C}^{q\times p},\\\nonumber \quad V=[v_1^T, \dots, v_q^T]^T=[\ell_1^TH(\mu_1), \dots,\ell_q^TH(\mu_q)]^T\in\mathbb{C}^{q\times m},
\end{align}
and a left part
\begin{align}\label{eq:leftData}
    \Lambda=\diag(\lambda_1,\dots,\lambda_k)\in\mathbb{C}^{k\times k}, \quad R=[r_1^T,\dots,r_k^T]^T\in\mathbb{C}^{m\times k},\\\nonumber\quad W=[w_1^T, \dots, w_k^T]^T=[H(\lambda_1)r_1, \dots,H(\lambda_q)r_k]^T\in\mathbb{C}^{p\times k}.
\end{align}
The aim is to find a rational function $H_r$ that fulfills the conditions in \eqref{eq:MIMOMM}. The data matrices defined in Equation~\eqref{eq:rightData} and Equation~\eqref{eq:leftData}, as well as the Loewner matrix $\mathbb{L}$ and its shifted version $\mathbb{L}_s$ defined via
\begin{equation}\label{eq:Loewner}
    \mathbb{L}:=\left(\frac{v_i^Tr_i-\ell_jw_j}{\mu_i-\lambda_j}\right)_{i=1,\dots,q}^{j=1,\dots,k}, \quad \mathbb{L}_s:=\left(\frac{\mu_iv_i^Tr_i-\lambda_j\ell_j^Tw_j}{\mu_i-\lambda_j}\right)_{i=1,\dots,q}^{j=1,\dots,k},
\end{equation}
are used to construct the unprocessed Loewner model $\{W,\mathbb{L},\mathbb{L}_s,V\}$. If the pencil $(\mathbb{L},\mathbb{L}_s)$ is regular, $H_r(s):=W(\mathbb{L}_s-s\mathbb{L})^{-1}V$ satisfies the interpolation condition in \eqref{eq:MIMOMM}. In many applications the pencil $(\mathbb{L},\mathbb{L}_s)$ is singular and a post-processing step is required. The dominant features of the data are extracted and redundancies removed. Commonly this is done by applying the singular value decomposition (SVD) to the augmented Loewner matrix
\begin{equation}\label{eq:LoewnerSVD}
    \begin{bmatrix}\mathbb{L} & \mathbb{L}_s\end{bmatrix} = Y\tilde{\Sigma}\tilde{X}^H, \quad \begin{bmatrix}\mathbb{L} \\ \mathbb{L}_s\end{bmatrix} = \tilde{Y}\Sigma X^H,
\end{equation}
with $\Sigma,\tilde{\Sigma}\in\mathbb{C}^{r\times r}$, $Y\in\mathbb{C}^{q\times r}$, $\tilde{Y}\in\mathbb{C}^{2q\times r}$, $X\in\mathbb{C}^{k\times r}$ and $\tilde{X}\in\mathbb{C}^{r\times 2k}$. The truncation index $r$ is chosen depending on the application and data size. The projected Loewner model is then given by
\begin{equation}\label{eq:LoewnerProjection}
   \quad A_r = -Y^T\mathbb{L}_sX, \quad B_r = Y^TV, \quad C_r = WX.
\end{equation}
The SVD provides optimal low-rank solutions, however, the full SVD has cubic complexity. An alternative way was proposed in \cite{Karachalios2021} and uses the CUR decomposition instead, which is less accurate, but has a lower asymptotic time complexity. Moreover, it has the benefit that the rows and columns of the decomposed matrices $\mathcal{C}$ and $\mathcal{R}$ are interpretable. In our case, this means that the dominant interpolation points $\lambda$ and $\mu$ are given directly by the columns and rows of $\mathcal{C}$ and $\mathcal{R}$.

There exists an iterative extension of the Loewner framework called AAA algorithm \cite{Antoulas1986}. Within this method the fitted rational approximants are expressed in a numerically stable way using a baricentric representation. In every iteration the next interpolation points are selected via a greedy method, i.e. interpolation is enforced at data points where the error was maximal. Recently a version of the algorithm enforcing real-valued and strictly proper rational approximants was proposed in \cite{Gosea2021}.

\section{Reduced order high fidelity mirror simulations}\label{sec:digital_twin}
In the process of designing a specific adaptive mirror there is the need to perform simulations that accurately represent the structural dynamics, i.e., realistic operative conditions. The simulations have to guarantee high fidelity results, but with an affordable computational load. As the size of the models is very large, model order reduction methods are required. 

\begin{figure}[ht]
\centering
  \includegraphics[width=1.\textwidth]{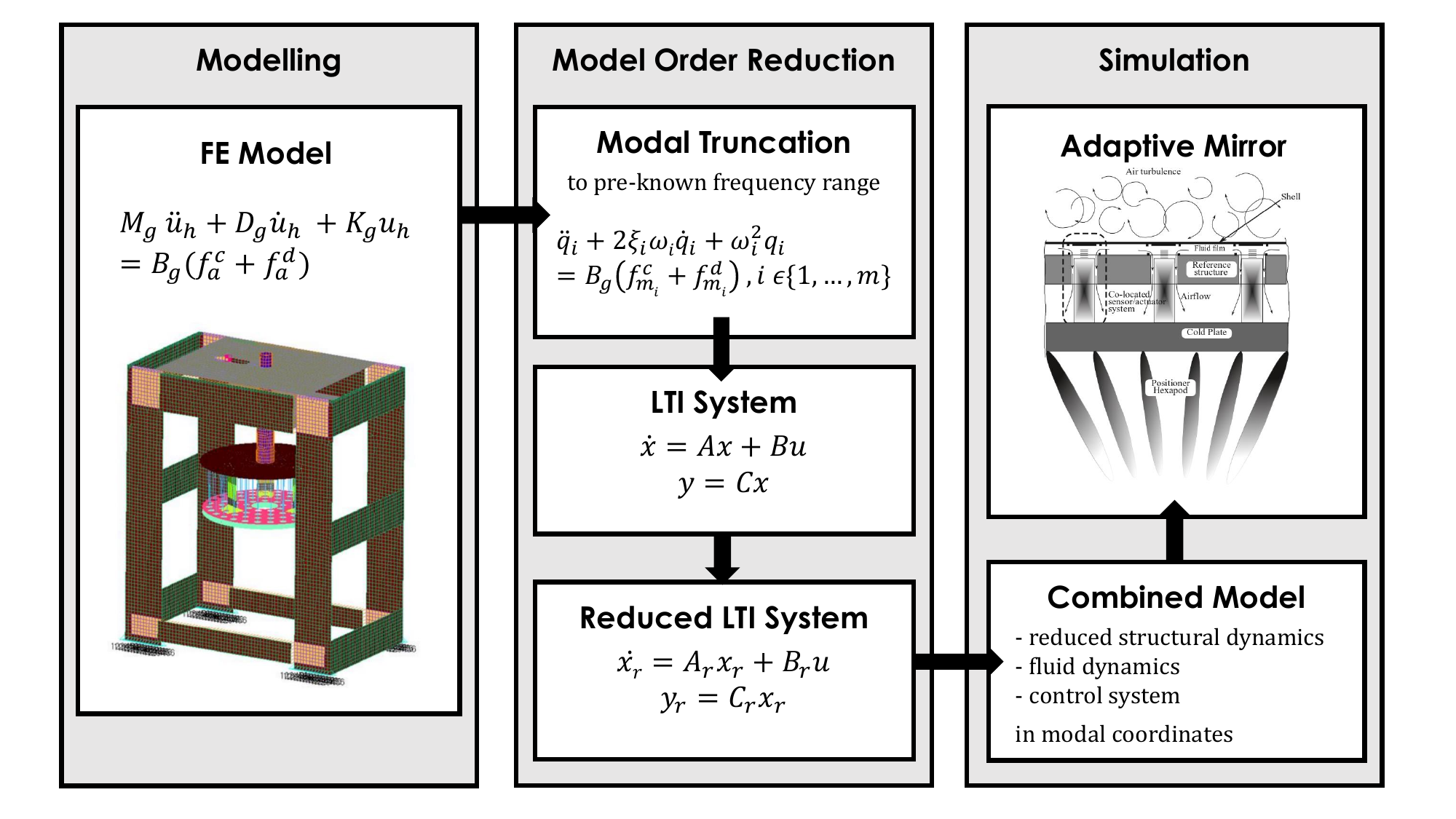}
  \caption{Graphical illustration of performing high fidelity adaptive mirror simulations with reduced order models.}
  \label{fig:romProcess}
\end{figure}

Figure~\ref{fig:romProcess} illustrates the process of performing high fidelity adaptive mirrors simulations with reduced order models. In a first step the accurate and complex FE models describing the deformable mirror, all the system supporting structure and in some cases even the full telescope are created. These models have been provided by the company A.D.S. International\footnote{\url{https://www.ads-int.com/}}. The FE models are then reduced by modal truncation to a predefined frequency range of interest for the certain adaptive mirror. As a second step of model order reduction, methods described in Section~\ref{sec:mor} are used to reduce the system size further. The resulting models are then combined with the rest of the system, i.e., fluid dynamics and control system, and the dynamic analysis of the mirror is performed. Especially, the second step of model order reduction makes it possible for Microgate to run their many simulations required in the adaptive mirror's design process in a feasible time frame.

In the upcoming sections the model order reduction and system simulations are described in more detail.

\subsection{Creating the reduced order mirror model}
The FE model is given in modal form \eqref{eq:modalDispl}. The first step of model reduction is applied by truncation to a given frequency range of the mirror as described in Section~\ref{sec:modalTruncatin}. Based on the resulting reduced system the LTI state space representation in modal form is set up as shown in Equation~\eqref{eq:modalLTI}. A second step of model order reduction is applied using either balanced truncation, rational Krylov subspace methods or the Loewner framework. More details on the implementation of the methods is given in the upcoming sections. Our code is mainly implemented in MATLAB. Computationally demanding parts are programmed in C. Parts of the C implementation use the Matrix Equation Sparse Solver (M.E.S.S) library \cite{SaaKB21} to efficiently deal with matrices, factorization and solving systems of equations.

\subsubsection{Balanced truncation}
Algorithm~\ref{alg:BT} shows the steps to perform the model order reduction with balanced truncation. In order to make the method feasible for the given system size, the modal approximation as shown in Equation~\eqref{eq:GramiansModal1} and Equation~\eqref{eq:GramiansModal2} is used. The modal controllability factor $w_{c_i}$ and observability factor $w_{o_i}$ and the Hankel singular values $\gamma_i$ are computed per mode $i=1,\dots,n$. The Hankel singular values are then ordered and all values below a certain threshold removed. As there is a one-to-one mapping between the Hankel singular values and the modes, the corresponding modes are discarded and in this way the system dimension reduced.

\begin{algorithm}
\caption{Balanced truncation in modal form}
\label{alg:BT}
\begin{algorithmic}[1]
\Function{ModalBT}{$B_{m},C_{m},\xi,\omega, t$}
    \For{each mode $i=1,\dots,n$}
    \State Compute the modal controllability factors $w_{c_i}$ via Equation~\eqref{eq:GramiansModal1}.
    \State Compute the modal observability factors $w_{o_i}$ via Equation~\eqref{eq:GramiansModal2}.
    \State
    \State Compute the Hankel singular values $\gamma_i$ using Equation~\eqref{eq:HankelModal}.
    \EndFor
    \State
    \State Order the Hankel singular values $\gamma_1\leq\gamma_2\dots\leq\gamma_n$.
    \State Discard all Hankel singular values below the given threshold $t$
    \State \quad $\gamma_1\leq\gamma_2\dots\leq\gamma_{j-1}$ where $\gamma_{j-1}\leq t<\gamma_j$.
    \State
    \State Compute $B_{m}^r,C_{m}^r,\xi^r$ and $\omega^r$ from the higher order model by discarding 
    \State all modes $i$ corresponding to the Hankel singular values $\gamma_1,\dots,\gamma_{j-1}$.
\EndFunction
\end{algorithmic}
\end{algorithm}

\subsubsection{Rational Krylov subspace methods}
Algorithm~\ref{alg:Krylov} illustrates the model order reduction procedure with Krylov subspace based methods. Our studies showed that the quality of these methods depends highly on the chosen interpolation points and that methods which use multiple expansion points and update them iteratively perform best. Hence, we consider here \ac{ITIA},\ac{ISTIA} and\ac{FISTIA}, see Section~\ref{sec:mor} for more details. In a first step of the algorithm the initial interpolation points are chosen. The choice of the points is important as the methods provide only local convergence. The eigenvalues of the state matrix would be a good initial guess. However, computing them for a large scale system is not feasible. Because our input FE model is given in modal form, we use their modal approximation, see e.g. \cite{Gawronski2004}. In general, one would choose the initial interpolation directions $b$ and $c$ as the corresponding left and right eigenvectors. Due to computational reasons we omit the computation and set $b$ and $c$ to vectors of ones. The projection matrices $V$ and $W$ are computed and used to obtain the reduced state space system $(A_r,B_r,C_r)$. For details on the computation of the projection matrices we refer to \cite{VANDOOREN2008,Vassal2011}. The interpolation points for the next iteration are chosen as the eigenvalues of the state matrix $\lambda(A_r)$ and the interpolation directions as the corresponding left and right eigenvectors. In this way the interpolation points and directions are updated in an iterative way until a maximum number of iterations is reached. Although the input to the reduced order methods is in modal coordinates, the Krylov subspace based methods produce a general state space representation. In order to study the system performance in an efficient way we transform these state space matrices into modal form.

\begin{algorithm}
\caption{Rational Krylov subspace method}
\label{alg:Krylov}
\begin{algorithmic}[1]
\Function{RationalKrylov}{$A_{m},B_{m},C_{m},maxIter$}
    \State
    \For{each mode $i=1,\dots,n$}
    \State Compute the interpolation points $s_i = \xi_i\omega_i \pm j\omega_i\sqrt{1-\xi^2_i}$.
    \State Set up the interpolation directions $b_i = [1,\dots,1], c_i = [1,\dots,1]$.
    \EndFor
    \State
    \For{$i=1,\dots,maxIter$}
    \State Compute the projection matrices $V$ and $W$.
    \State Compute the reduced system via Equation~\eqref{eq:Petrov-Galerkin}.
    \State Compute eigenvalues $\lambda$ and left and right eigenvectors $U_\ell,U_r$ of $A_r$.
    \State Extract the new interpolation points $s=-\lambda$.
    \State Extract the new interpolation directions $b = U_{\ell}$, $c=U_{r}$.
    \EndFor
    \State
    \State Verify that the reduced system $(A_r,B_r,C_r)$ is stable.
    \State Transform $(A_r,B_r,C_r)$ into modal coordinates $(A_m^r,B_m^r,C_m^r)$.
\EndFunction
\end{algorithmic}
\end{algorithm}

\subsubsection{Loewner framework}
Algorithm~\ref{alg:Loewner} shows how to create a reduced order model using the Loewner framework. In a first step $N$ logarithmically spaced interpolation points are created over the frequency range of interest. The reason for choosing logarithmically distributed points is that in adaptive mirror control it is more important to match the lower frequencies. Hence, having a dense sampling grid in the beginning and a more sparse distribution towards the end of the frequency range is more appropriate for our application. The initial interpolation directions $b$ and $c$ are chosen as random values. Using the input state space system in modal coordinates $(A_m,B_m,C_m)$, the transfer function $H_m(s)$ is constructed and the right and left data set is computed as shown in Equation~\eqref{eq:rightData} and Equation~\eqref{eq:leftData}. Note that as the state space system is given in modal coordinates the evaluation of the transfer function and thus the computation of the input-output pairs for the data set is much faster. For more details we refer to \cite{Gawronski1990}. The data set is then used to compute the Loewner matrix $\mathbb{L}$ and its shifted version $\mathbb{L}_s$ using Equation~\eqref{eq:Loewner}. The constructed complex representation $(W,\mathbb{L},\mathbb{L}_s,V)$ of an underlying dynamical system is transformed into a real model $(W^r,\mathbb{L}^r,\mathbb{L}_s^r,V^r)$, see e.g. \cite{Karachalios2021}. To identify dominant subsets of the data on which interpolation is enforced we apply the SVD to the Loewner matrices using Equation~\eqref{eq:LoewnerSVD}. Note that the CUR decomposition would provide a faster way of achieving this, however, it is only an approximation and performed worse in our experiments. The reduced order model is then obtained through projection using Equation~\eqref{eq:LoewnerProjection}. Similar as for the Krylov subspace based method we transform the system into modal coordinates as a last step.

\begin{algorithm}
\caption{Loewner framework with SVD}
\label{alg:Loewner}
\begin{algorithmic}[1]
\Function{LoewnerSVD}{$A_{m},B_{m},C_{m},N$}
    \State Compute the interpolation points $s = \text{logspace}(-1,3,N)$.
    \State Set up interpolation directions $b = \text{rand}(n)$ and $c = \text{rand}(n)$.
        \State
    \State Compute the transfer function $H_m$.
    \State Construct the right data set $M,L$ and $V$ using Equation~\eqref{eq:rightData}.
    \State Construct the left data set $\Lambda, R, W$ via Equation~\eqref{eq:leftData}.
    \State
    \State Compute the Loewner pencil $(\mathbb{L},\mathbb{L}_s)$ via Equation~\eqref{eq:Loewner}.
    \State Transform the complex system into a real system $(W^r,\mathbb{L}^r,\mathbb{L}_s^r,V^r)$.
    \State
    \State Compute the SVD of the Loewner matrix using Equation~\eqref{eq:LoewnerSVD}.
    \State Compute the reduced model $(A_r,B_r,C_r)$ via Equation~\eqref{eq:LoewnerProjection}.
    \State
    \State Verify that the reduced system $(A_r,B_r,C_r)$ is stable.
    \State Transform $(A_r,B_r,C_r)$ into modal coordinates $(A_m^r,B_m^r,C_m^r)$.
\EndFunction
\end{algorithmic}
\end{algorithm}

In our experiments we also considered the AAA algorithm \cite{Antoulas1986}, which is an adaptive and iterative extension of the Loewner framework that chooses automatically the reduction size for a given accuracy. However, for our simulation setting the results have been worse compared to the Loewner framework and thus we omit more details on this method.

\subsection{System simulation} \label{subSec:SysSim}
The reduced order model of the system structural dynamics are combined with the remaining system and exploited to run dynamic analysis. Different numerical tools have been developed for this purpose:
\begin{itemize}
    \item dynamic simulation to evaluate the system behavior in time, this simulation is mainly focused on the analysis of the system performance and stability;
    \item frequency response to have a characterization in the frequency domain, very useful to understand possible critical frequencies in the dynamic response of the system, but also to provide a measurement of system robustness;
    \item root locus analysis to obtain a description of the system in the Laplace domain, having a characterization of the system stability and stability sensitivity, i.e. robustness, with respect to some specific design parameters.    
\end{itemize}
The three numerical tools above mentioned can be used all together to provide a comprehensive overview of the system behavior in terms of both stability and performance. The accuracy of the system description provided by these different tools is usually not equivalent and the fidelity level of each one can be tuned to find the best compromise between accuracy of the results and computational time and resources necessary to perform the analysis. 

In the framework of this paper Microgate's mirror time simulator has been upgraded to a new version, based on a multi-threaded parallel C++ implementation. This new code can provide the most reliable system multiphysics simulation currently available and contains all the features of the previous implementation written in C language \cite{Manetti2010,Manetti2012}. The new simulator has the additional relevant capability to take into account not only the structural dynamics of the deformable shell, but of the whole system supporting structure, potentially up to the telescope one. The order of the structural dynamic description must be reduced as much as possible to allow fully exploiting this new feature of the simulator, in order to prevent that the simulation time and/or the amount of computational resources become a show stopper for the analysis. The codes computational efficiency has been further improved by using the blaze\footnote{\url{https://github.com/parsa/blaze}} library, which is an open-source, high-performance, C++ math library for dense and sparse arithmetic, and OpenMP\footnote{\url{https://www.openmp.org/}} for parallelization.
 
\section{Numerical analysis}\label{sec:numerical_simuls}
\begin{figure}[ht]
\centering
  \includegraphics[width=0.8\textwidth]{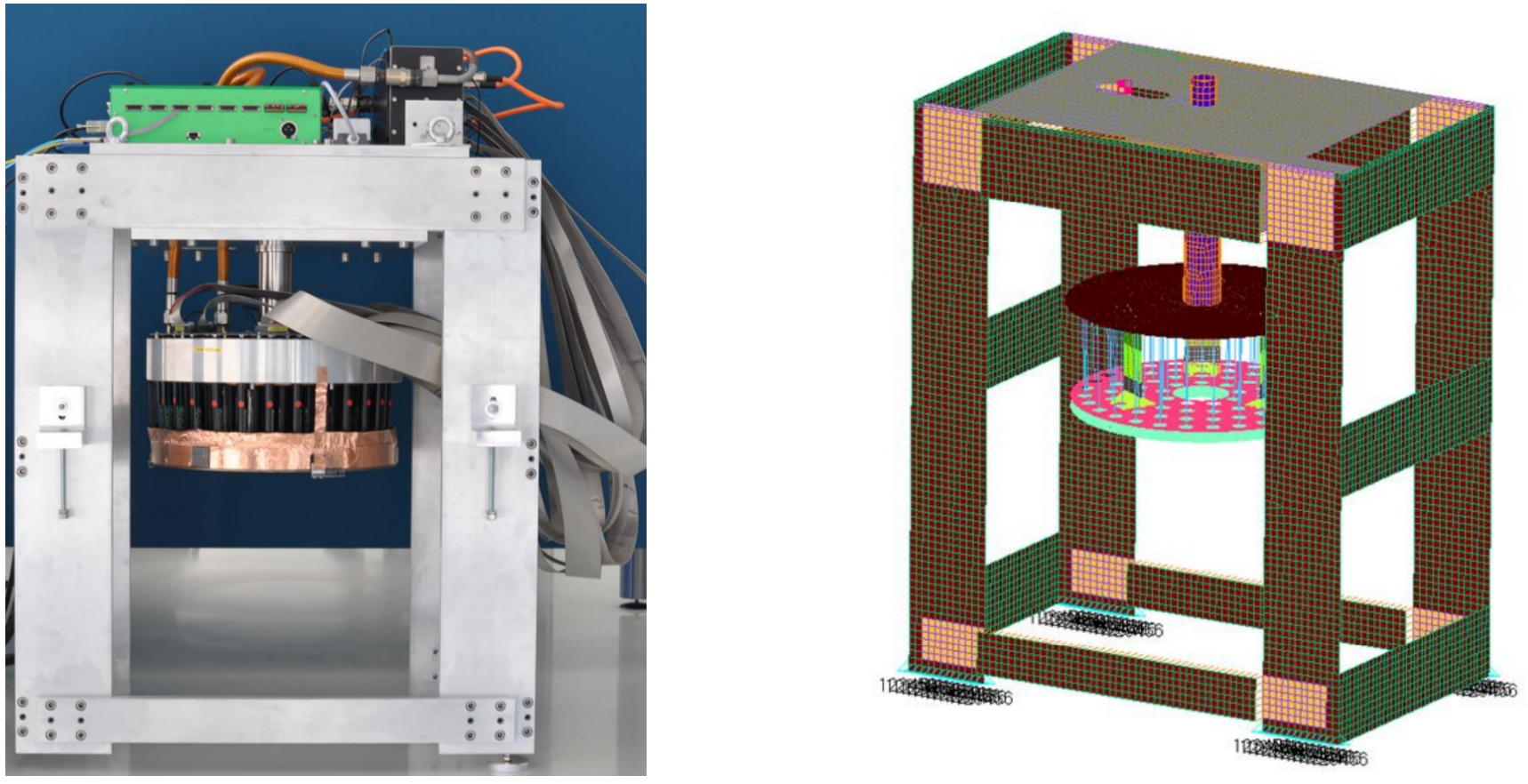}
  \caption{Picture (left) and FE model (right) of the GMT P72 \cite{Gallieni2020}.}
  \label{fig:GMTP72}
\end{figure}

To validate our developments we perform numerical analysis for the model of the P72 prototype of the GMT adaptive secondary mirror system \cite{Gallieni2020}. The GMT is currently under construction in the Atacama desert in Chile and will become one of the new ELTs \cite{Gallieni2020}. The prototype has $72$ actuators and a diameter of $354$ mm, featuring the four innermost rings of actuators of the on-axis adaptive secondary mirror segment of the GMT, see Figure~\ref{fig:GMTP72}. The shell is $2$ mm thick and made of Zerodur, the lateral flexures connecting the shell to the \ac{RB} are the same of the final unit and also the VCM actuators are exactly the final ones. The RB is made in Zerodur, while the cold plate, where the electronics is mounted, is in aluminum. The cooling system is based on the direct expansion gas cooling concept recently introduced in this application field\cite{Pescoller2017}.  All the simulators mentioned in Section~\ref{subSec:SysSim} exploit the very same FE model, shown in Figure~\ref{fig:GMTP72}, to capture the system structural behavior, through the use of eigenmodes.

\subsection{Dynamic model order reduction}
The FE model of the GMT P72 prototype is given in modal form and we truncate it to the pre-known frequency range $0-6$kHz. Based on the resulting reduced system we set up the LTI state space model and compare \ac{BT}, \ac{ITIA},\ac{ISTIA} and the \ac{LF}. The \ac{HO} model after modal truncation has a state matrix with dimension $1672\times1672$. For \ac{ISTIA} we use the controllability Gramian as one of the projection matrices. This choice provided slightly better results than using the observability Gramian.

\begin{figure}[ht]
\centering
\begin{subfigure}{.48\textwidth}
  \centering
    \includegraphics[width=1.\textwidth]{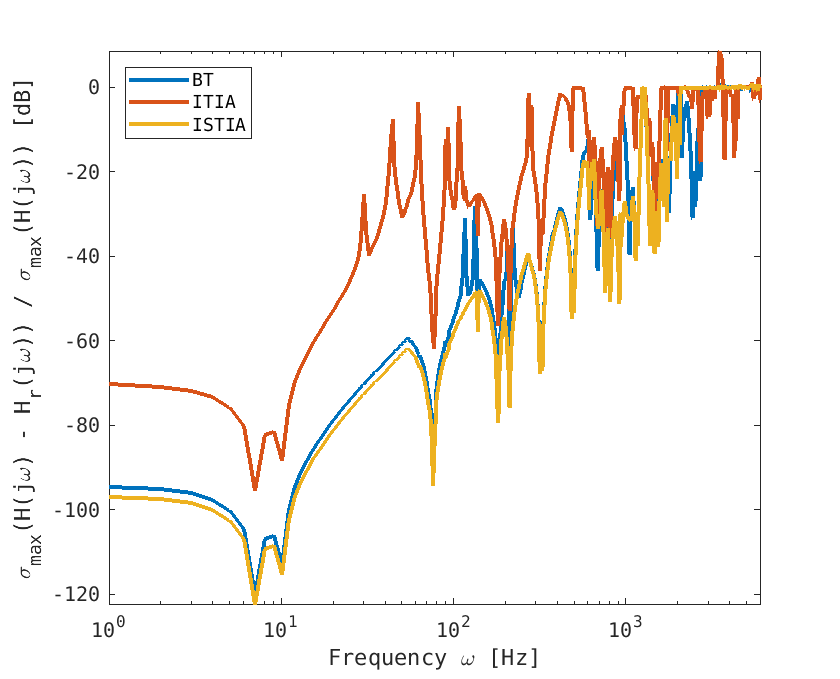}
  \caption{}
\end{subfigure}
\begin{subfigure}{.5\textwidth}
  \centering
  \includegraphics[width=1.\textwidth]{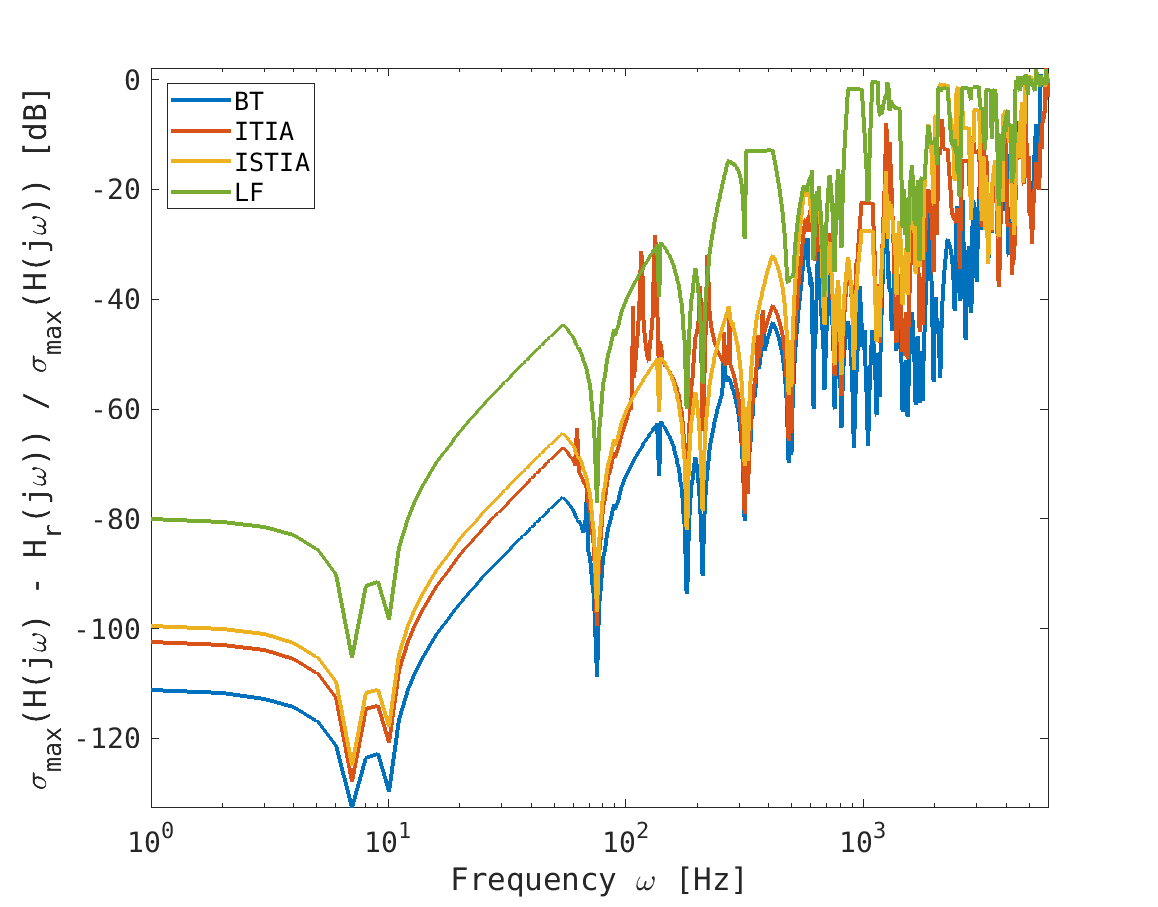}
  \caption{}
\end{subfigure}
  \caption{Logarithmic plot of the relative $\mathcal{H}_\infty$ error in dB between the \ac{HO} and the reduced order transfer functions with reduced state matrices of dimension $144\times 144$ (left) and $330 \times 330$ (right) for \ac{BT} (blue), \ac{ITIA} (orange), \ac{ISTIA} (yellow) and \ac{LF} (green).}
  \label{fig:errorTF}
\end{figure}

Figure~\ref{fig:errorTF} shows a logarithmic plot of the relative $\mathcal{H}_\infty$ error of the transfer functions between the \ac{HO} and the reduced order models. In plot (a) we show the model reduction error for a reduced state matrix of dimension $144\times 144$ and in (b) for a size of $330 \times 330$. The reason for choosing $144$ as lower limit is that the P72 consists of $72$ degrees of freedom resulting in a state matrix of $144\times 144$. Hence, this is the lowest order allowed to be able to statically set the mirror deformation over $72$ actuation points. During our studies we obtained the best trade-off between quality and speed with a state matrix of size $330 \times 330$. We compare here \ac{BT} marked in blue, \ac{ITIA} in orange, \ac{ISTIA} in yellow and \ac{LF} in green. Note that for this setting \ac{ISTIA} and \ac{FISTIA} performed very similar, hence, we omit \ac{FISTIA}. The reason for this might be that the model was already truncated to the desired frequency range by modal truncation. In Table~\ref{tab:meanErrorTF} we list the mean of the relative error over the whole frequency range. For both simulated cases \ac{BT} and the Krylov subspace methods performed very similar. With \ac{LF} we were not able to produce an appropriate reduced order model with a size of $144 \times 144$, hence, we do not show it in Figure~\ref{fig:errorTF}(a). For a reduced system size of $330 \times 330$ \ac{LF} yields the highest $\mathcal{H}_\infty$ error. This might be caused by not optimally chosen interpolation points. For the other methods an increased reduction size provides a smaller mean relative error.

\begin{center}
\begin{table}[ht]
\centering
\def\arraystretch{1.3}
\pgfplotstabletypeset
[
    col sep=&,	
    row sep=\\,	
    columns={a,b,c,d,e},  
    columns/a/.style={
        column name={\color{black}\bfseries{Reduction size}},
        column type={|R{0.2\linewidth}|},
        string type
    },
    columns/b/.style={
        column name={\color{black}\bfseries{BT}},
        column type={|C{0.1\linewidth}|},
        string type
    },   
    columns/c/.style={
        column name={\color{black}\bfseries{ITIA}},
        column type={|C{0.1\linewidth}|},
        string type
    },
    columns/d/.style={
        column name={\color{black}\bfseries{ISTIA}},
        column type={|C{0.1\linewidth}|},
        string type
    },
    columns/e/.style={
        column name={\color{black}\bfseries{LF}},
        column type={|C{0.1\linewidth}|},
        string type
    },
    every head row/.style={
        before row={\hline
        & \multicolumn{4}{c|}{\bfseries Mean relative $\mathcal{H}_\infty$ error}\\},
        after row={\hline},
    },  
    every last row/.style={after row=\hline}
]
{
    a & b & c & d & e\\
    $144\times 144$ & $0.627$ & $0.768$ & $0.691$ & $-$\\
    $330\times 330$ & $0.158$ & $0.153$ & $0.368$ &
    $0.556$\\
} 
\caption{Mean of the relative $\mathcal{H}_\infty$ error as shown in Figure~\ref{fig:errorTF}. The mean is calculated over the desired frequency interval $0-6$kHz for \ac{BT} and the Krylovs subspace based methods \ac{ITIA} and \ac{ISTIA} and \ac{LF}.}
\label{tab:meanErrorTF}
\end{table}
\end{center}

Throughout the simulations we observed that the results of the Krylov subspace based methods and \ac{LF} highly depend on the choice of the initial interpolation points and directions. Note that for \ac{ITIA} and \ac{ISTIA} we use the modal approximation of the eigenvalues of the \ac{HO} state matrix. This offers an easy way of running the algorithms without tuning it by hand. For  \ac{LF} we use $6000$ logarithmically spaced points.

In terms of run-time \ac{BT} clearly outperforms the other methods, because computing the modal approximation of the Hankel singular values is very fast. For the Krylov subspace based methods the run-time depends mainly on the number of iterations and the reduction size, since in every iteration the eigenvalues and eigenvectors of the reduced order model have to be computed. In our simulations we obtained convergence within less than $50$ iterations. For the data-driven \ac{LF} we observed that a fairly high amount of data points is required to obtain an appropriate reduced order model, which increases the computational time as well as the memory requirements significantly.

Further analysis showed that all reduced order methods are not able to represent $72$ independent degrees of freedom with a state matrix of $144\times 144$. Hence, this reduction size is too low to represent properly the system behavior and we continue our analysis with a reduced order model of size $330\times 330$. The performance of \ac{ISTIA} and \ac{ITIA} is very similar for the P72 mirror, hence, we focus for the upcoming analysis on \ac{ITIA}.

\subsection{Performance of the system}
In this section the reduced order models representing the system structural dynamics are exploited to simulate the full system behavior and a comparison of the results with the system using the original full set of eigenmodes is done. The whole system modeling can be summarized by the following list:
\begin{itemize}
    \item The structural dynamics of the system, including: the deformable mirror, the reference body, the cold plate and the P72 supporting structure.
    \item The fluid dynamic modeling of the air trapped between the deformable mirror and the reference body.
    \item The inner control loop, controlling the mirror shape, including: the deformable mirror control law, the voice coil motors dynamics, the capacitive sensors dynamics, the digital and analog signals modeling.
\end{itemize}
The optical loop, which is responsible to generate the mirror commands, is not taken into account within the present analysis, i.e, the mirror is simulated as a stand-alone component of the AO system.

\subsubsection{Stability}
\begin{figure}[ht]
  \centering
    \includegraphics[width=1.\linewidth]{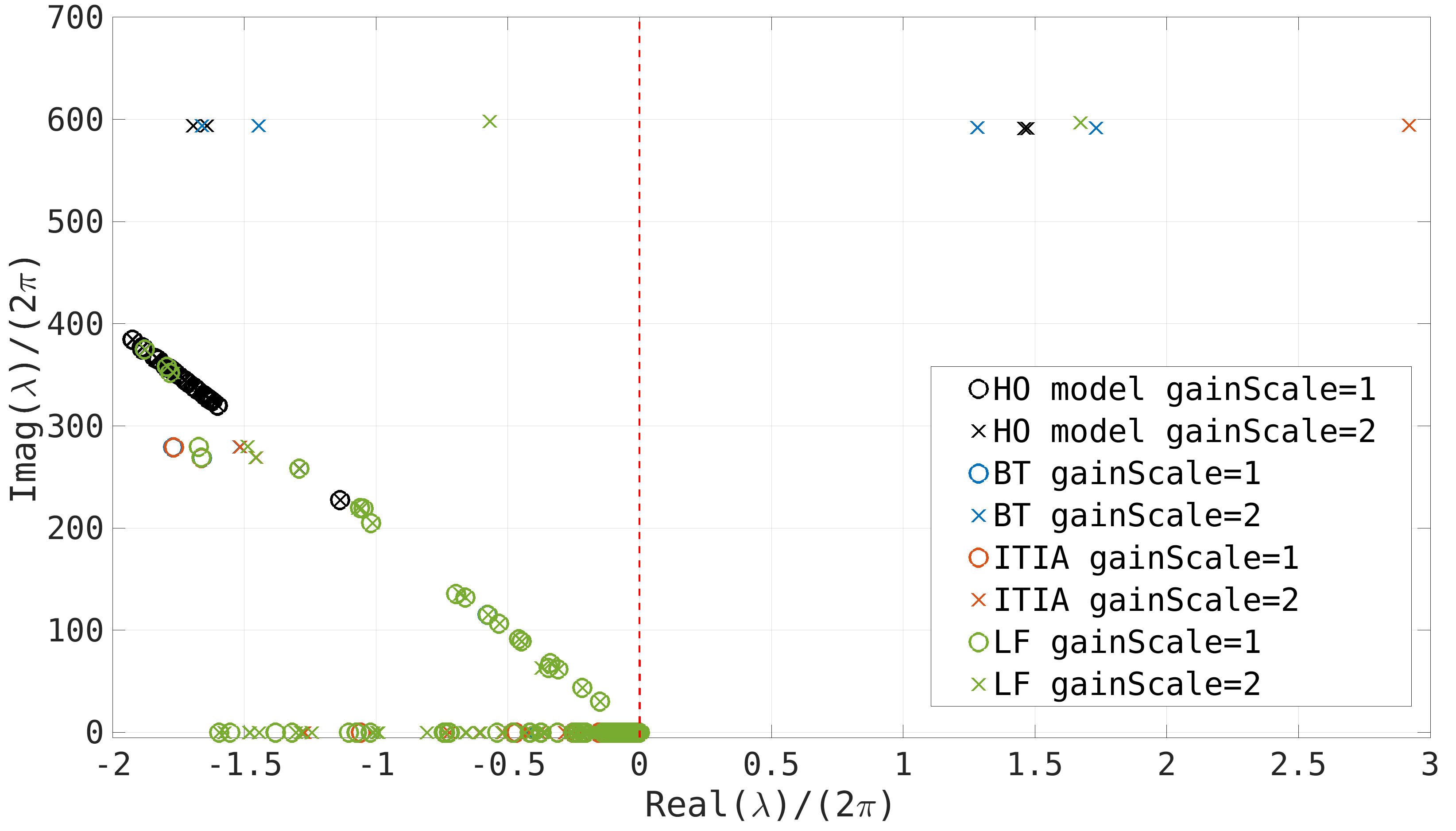}
\caption{Root locus plot of the \ac{HO} model in black and the reduced order model with size $330\times 330$ with \ac{BT} in blue, \ac{ITIA} in orange and \ac{LF} in green. The gain scaling is varied between $1$ and $2$. We show only the important section around the real $0$ axis (dashed red line). Moreover, we show only the poles with positive imaginary part and not their conjugate complex counterpart. Circles indicate stable systems, whereas crosses indicate unstable systems.}
\label{fig:rootLocus}
\end{figure}

To analyze the stability of the control system we use root locus plots, i.e., we plot the poles of the transfer function in the Laplace domain. These plots allow to study the sensitivity of the system against certain parameters. Here we consider the feedback gain with a varying scaling factor. We use a circle to indicate a stable system, whereas crosses for unstable systems. In terms of model order reduction, it is important that the reduced model does not indicate a stable system when the original model was not stable or vice versa. Figure~\ref{fig:rootLocus} shows the root locus plots of the higher order model (black) and of reduced order models of size $330\times 330$, using \ac{BT} (blue), \ac{ITIA} (orange) and \ac{LF} (green). Note that only the important section around the $0$ real axis is plotted, just for one half of the complex conjugate poles characterized by a positive imaginary part. We observe that all models indicate a stable system with gain scaling $1$ and an unstable system for a scaling of $2$. The critical poles, i.e. the ones having positive real part, slightly change for different methods. Our simulations showed that for smaller reduction sizes the reduced model may lead to stable systems where the original model was unstable.

\subsubsection{Input-output behavior}
\begin{figure}[ht]
\centering
    \includegraphics[width=1.\textwidth]{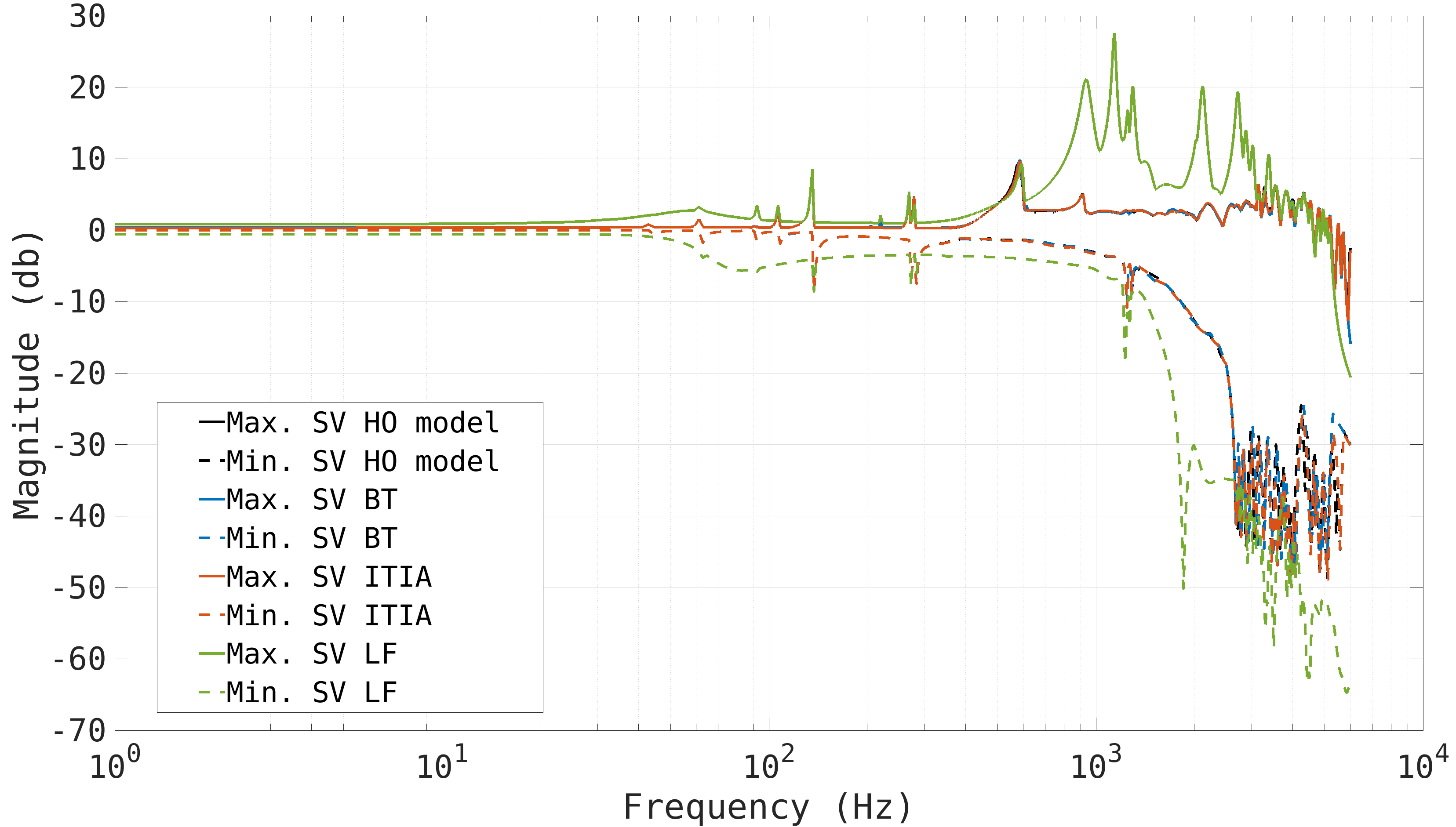}
  \caption{Logarithmic plot of the minimum (dashed) and maximum (solid) singular values in dB of the high and reduced order transfer functions in the frequency interval $0-6$ kHz. We compare the \ac{HO} model in black and the reduced order models with \ac{BT} in blue, \ac{ITIA} in orange and \ac{LF} in green.}
  \label{fig:sv330}
\end{figure}

To assess the difference in the input-output behavior of the full system, we study the minimum and maximum singular values of the complementary sensitivity transfer function matrix, defined to have the deformable mirror commands as inputs and the mirror deformation at the actuation points as output. The frequency response of the complementary sensitivity is commonly used to asses the tracking capabilities of the closed-loop system as function of the command frequency content. Figure~\ref{fig:sv330} shows a logarithmic plot of the minimum (dashed) and maximum (solid) singular values in dB for the \ac{HO} system in black and the reduced systems with \ac{BT} in blue,\ac{ITIA} in orange and \ac{LF} in green. We observe that the reduced order models obtained with \ac{BT} and \ac{ITIA} yield very similar maximum and minimum singular values compared to the original system. In contrast, the singular values produced by  \ac{LF} are significantly different.

\subsection{Mirror simulator step response}
\begin{figure}[ht!]
\begin{subfigure}{1.0\textwidth}
  \centering
  \includegraphics[width=1.\linewidth]{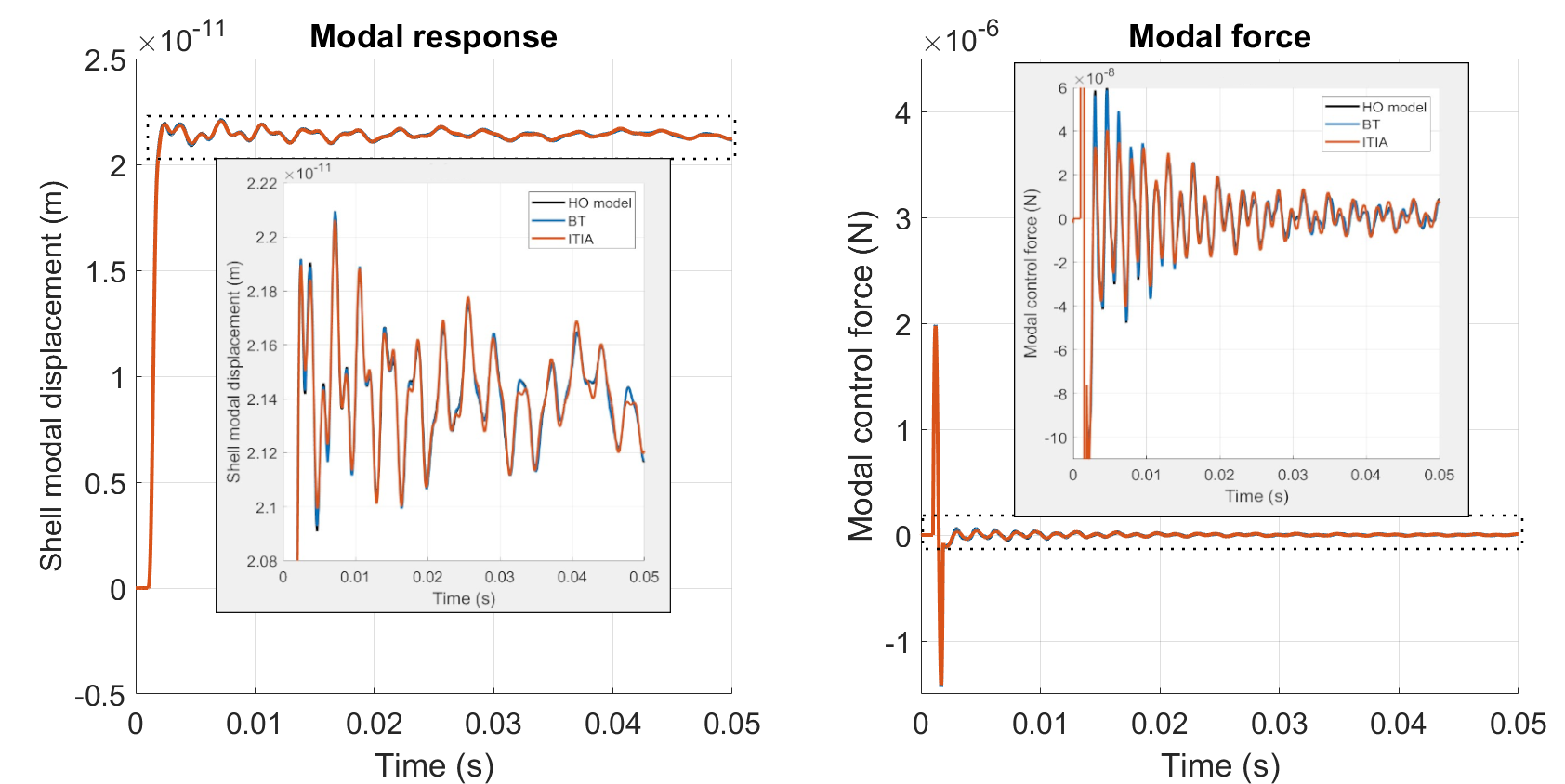}
  \caption{Step response for a tilt command. The left plot shows the shell displacement over time, whereas on the right the modal control force is shown.}
\end{subfigure}
\begin{subfigure}{1.\textwidth}
  \centering
    \includegraphics[width=1.\linewidth]{stepResponseTrefoil_new.png}
  \caption{Step response for a trefoil command. The left plot shows the shell displacement over time, whereas on the right the modal control force is shown.}
\end{subfigure}
\caption{Plot of the step response for a tilt command (a) and a trefoil command (b) for the \ac{HO} model in black and the reduced models with \ac{BT} (BT) in  blue and \ac{ITIA} in orange. In addition, a zoom around the y-axis is shown.}
\label{fig:TimeSimul}
\end{figure}

The last system analysis uses the reduced order models as input to the C++ mirror simulator and studies the simulation over time of the control system response to a certain mirror deformation command. In Figure~\ref{fig:TimeSimul} we show the step response plots to a tilt (a) and trefoil (b) command. We omit the reduced model produced by \ac{LF} as it lead to too large forces for the voice coil motor and thus the simulation was stopped. The plots on the left represent the shell displacement, whereas the ones on the right show the modal control force. We provide a zoom around the y-axis in order to be able to see the performance of the reduced order methods \ac{BT} in blue and \ac{ITIA} in orange compared to the \ac{HO} model in black. We observe that for both mirror commands all methods provide good results. However, the step response obtained with \ac{BT} is more in line with that of the original model, especially for the trefoil command case.

\section{Conclusion}\label{sec:conclusion}
In the design phase of an adaptive mirror, numerical simulations are crucial to evaluate the system design compliance. The FE models for the adaptive mirrors of ELTs have a high complexity, hence, model order reduction techniques that facilitate the computationally efficient analysis are required. In this paper we present a framework for performing high fidelity control system simulations within a reasonable time frame. In a preprocessing step a reduced order model is created, which is then used to simulate the full adaptive mirror. We perform a feasibility study of different reduced order methods exploiting the numerical model of the P72 prototype of the GMT secondary mirrors. Starting with the FE model in modal coordinates several model order reduction algorithms are applied. The quality is analyzed by means of the $\mathcal{H}_\infty$ error of the transfer functions. The reduced order model of the structural dynamics is combined with the remaining system modeling and its performance is analyzed regarding accuracy, stability and robustness. \ac{BT} and \ac{ITIA} yield both very similar stable and accurate models when reducing to a state matrix of size $330\times 330$ or larger. In terms of computational load, \ac{BT} with modal approximation is significantly faster than the others. The data-driven \ac{LF} was only able to produce an appropriate reduced model when considering a larger reduction size. The implemented framework for performing highly accurate simulation with reduced order models together with the developed numerical analysis tools allow to improve the design process of adaptive mirrors and the relevant research and development activity at Microgate. Although the framework was tested here on a relatively small prototype the developments can be directly applied to larger systems and will allow simulations within a reasonable computational time.

\section*{Acknowledgement}
This research was funded in part by the Austrian Science Fund (FWF) SFB 10.55776/F68 ``Tomography Across the Scales'', project F6805-N36 (Tomography in Astronomy). For open access purposes, the authors have applied a CC BY public copyright license to any author-accepted manuscript version arising from this submission. Furthermore, the authors were supported by the Austrian Research Promotion Agency (FFG) project number FO999888133, as well as the NVIDIA Corporation Academic Hardware Grant Program. LW is partially supported by the State of Upper Austria.

\bibliographystyle{plain}
\bibliography{mybib}

\end{document}